\documentclass[useAMS,usenatbib]{mn2e}

\usepackage{epsfig}
\usepackage{amsmath}
\usepackage{amssymb}
\usepackage{natbib}
\usepackage{threeparttable} 
\usepackage{subfigure}

\def\simmore{\mathbin{\lower 3pt\hbox
     {$\rlap{\raise 5pt\hbox{$\char'076$}}\mathchar"7218$}}}   

\usepackage{graphicx}

\title[The cooling phase of Type-I X-ray bursts in 4U 1636--53]{The cooling phase of Type-I X-ray bursts in 4U 1636--53}

\author[G. Zhang et al.]{Guobao Zhang$^{1}$\thanks{E-mail:
zhang@astro.rug.nl}, Mariano M\'endez$^{1}$ and Diego Altamirano$^{2}$\\
$^{1}$Kapteyn Astronomical Institute, University of Groningen, P.O. BOX
800, 9700 AV Groningen, The Netherlands\\
$^{2}$Astronomical Institute, ``Anton Pannekoek'', University of Amsterdam, Science Park 904, 1098 XH Amsterdam, The Netherlands\\}

\begin{document}


\maketitle

\label{firstpage}

\date{Accepted. Received; in original form}

\begin{abstract}
Time-resolved spectra during the cooling phase of thermonuclear X-ray
bursts in low-mass X-ray binaries (LMXBs) can be used to measure the
radii and masses of neutron stars. We analyzed $\sim$ 300 bursts of the
LMXB 4U 1636--53 using data from the Rossi X-ray Timing Explorer. We
divided the bursts in three groups, photospheric radius expansion (PRE),
hard non-PRE and soft non-PRE bursts, based on the properties of the
bursts and the state of the source at the time of the burst. For the
three types of bursts, we found that the average relation between the
bolometric flux and the temperature during the cooling phase of the
bursts is significantly different from the canonical $F \propto T^4$
relation that is expected if the apparent emitting area on the surface
of the neutron star remains constant as the flux decreases during the
decay of the bursts. We also found that a single power law cannot fit
the average flux-temperature relation for any of the three types of
bursts, and that the flux-temperature relation for the three types of
bursts is significantly different. Finally, for the three types of
bursts, the temperature distribution at different flux levels during the
decay of the bursts is significantly different. From the above we
conclude that hard non-PRE bursts ignite in a hydrogen-rich atmosphere,
whereas for soft non-PRE and PRE bursts the fuel is helium-rich. We
further conclude that the metal abundance in the neutron star atmosphere
decreases as the bursts decay, probably because the heavy elements sink
faster in the atmosphere than H and He.

\end{abstract}


\section{introduction}
\label{introduction}

Thermonuclear, type-I, X-ray bursts are due to unstable burning of H and He on the surface
of accreting neutron stars in low-mass X-ray binaries (LMXBs). During these 
bursts, the observed X-ray intensity first sharply increases, typically by a factor of
$\sim 10$ in about 0.5$-$5 seconds, and after that it decreases more or less  
exponentially within 10$-$100 seconds \citep[e.g.,][]{lewin93, Strohmayer03, galloway}.
The total energy emitted by an X-ray burst is typically $\sim 10^{39}$ ergs.
Some X-ray bursts are strong enough to lift up the outer layers of the star.
During these so-called photospheric radius expansion (PRE) bursts \citep{Basinska}, 
the radiation flux that emerges from the stellar surface is limited by the Eddington flux. 
The neutron-star surface origin is supported by the fact that the inferred emission area 
during a burst matches the expected surface area of a neutron star, assuming that the 
thermonuclear flash expands to cover the entire star during the radius expansion and 
cooling phases of the burst. The emission area can be estimated from the fitting of 
the energy spectra during the cooling tails of bursts. There are a
number of theoretical and observational arguments that support this assumption 
\citep[see, e.g. ][]{Fryxel82, Bildsten95, Spitkovsky02, Strohmayer03}.

One of the best studied sources of X-ray bursts is the LMXB 4U
1636--53. Also known as V801 Ara, 4U 1636--53 was discovered with OSO-8
\citep{Swank}, and was subsequently studied in great detail using
observations with SAS-3, Hakucho, Tenma, and EXOSAT \citep[see][for a
review]{lewin87}. The orbital period of this binary system is 3.8 hr
\citep{van Paradijs90}, and the spin period of the neutron star is 581
Hz \citep{Strohmayer98a,Strohmayer98b}. Using EXOSAT, \cite{Damen89}
detected 60 bursts from this source between 1983 and 1986; from
observations with the Rossi X-ray Timing Explorer (RXTE), 172 bursts
were detected up to 2007 June 3 \citep{galloway}, and more than 250
bursts including data taken after that date \citep{zhang gb}.
Most of these X-ray bursts have standard, single-peaked, fast rising
and exponentially decaying light curves. Some single-peaked bursts 
show photospheric radius expansion,  while a few other  bursts from 
4U 1636--53 show multi-peaked light curves, although these are not 
PRE bursts \citep{maurer,Bhattacharyyaa,zhang gb}.

In this paper, we compare the temperature and emission-area distributions in PRE
non-PRE and double-peaked bursts during their cooling phase.

\section{Observation and data analysis}
\label{data}

We analyzed all data available from the RXTE Proportional Counter Array
(PCA) of 4U 1636--53 as of May 2010. The PCA consists of an array
of five collimated proportional counter units (PCUs) operating in the
2$-$60 keV range. We produced 1-s light curves from the Standard1 data 
(0.125-s time resolution with no energy resolution) and searched for X-ray
bursts in the light curves. We used the 16-s time-resolution Standard-2 data to 
calculate X-ray colours of the source, as described in \cite{zhang gb}. 
Hard and soft colours are defined as the $9.7-16.0/6.0-9.7$ keV and 
$3.5-6.0/2.0-3.5$ keV count rate ratios, respectively, and intensity 
as the $2.0-16.0$ keV count rate. The colour-colour diagram (CD) of all observations
of 4U 1636--53 is shown in Figure \ref{fig: CCD}. The position of 
the source on the diagram is parameterized by the length of the solid curve 
$S_{\rm a}$ \citep[see, e.g. ][]{mendez99}. The $S_{\rm a}$ length is normalized to the distance 
between $S_{\rm a}=1$ at the top right vertex, and $S_{\rm a}=2$ at the bottom left 
vertex of the CD. Similar to the $S_{\rm z}$ length in Z sources \citep{Vrtilek},
$S_{\rm a}$ is considered to be a function of mass accretion rate \citep{Hasinger89}.
 
In order to study the bursts in detail, we analyzed the PCA Event data,
E\_125us\_64M\_0\_1s, in which each individual photon is time tagged at
a $\sim 122$ $\mu$s time resolution in 64 energy channels between
$2-60$ keV. We used all PCUs that were operating at the time of the
X-ray burst to produce 0.5-s resolution light curves in the full PCA band. 
For the time-resolved spectral analysis of the
bursts we extracted spectra in 64 channels every 0.5 s from the Event
data of all available PCUs. We corrected each spectrum for dead time using 
the methods supplied by the RXTE team.  We generated the instrument response matrix 
files for each observation, and we fitted the spectra using XSPEC
version 12.4.0. Because of the short exposure, in this case the
statistical errors dominate, and therefore we did not add any
systematic error to the spectra. We restricted the spectral fits to the
energy range $3.0 -20.0$ keV. For each burst we extracted the spectrum from the persistent data
just before (or after) the burst to use as background in our fits; 
this approach, used to obtain the net emission of a burst, is well established as a 
standard procedure in X-ray burst analysis \citep[e.g.][]{Kuulkers02}.
We fitted the time-resolved net burst spectra with a single-temperature
blackbody model (bbodyrad in XSPEC), as generally burst spectra are
well fitted by a blackbody. During the fitting, we kept the hydrogen
column density $N_{\rm H}$ fixed at $0.36\times10^{22} $cm$^{-2}$
\citep{Pandel08}, and to calculate the radius of the emitting area in
km, we assumed a distance of 5.95 kpc \citep{fiocchi}. The model
provides the blackbody colour temperature ($T_{\rm bb}$) and a
normalization proportional to the square of the blackbody radius
($R_{\rm bb}$) of the burst emission surface, and allows us to
estimate the bolometric flux as a function of time.

We note that this procedure fails if the
blackbody emission during the burst comes from the same source that
produces the blackbody emission seen in the persistent emission, since
the difference between two blackbody spectra is not a blackbody
\citep{van Paradijs}. This effect is significant only when the net
burst emission is small, and therefore problems may arise only at
the start and tail of the burst, when the net burst emission is comparable
to the underlying persistent emission \cite[see the discussion in][]{Kuulkers02}. 

To test whether fitting the net spectrum of the bursts affects our results, we proceeded as follows:
For a subsample of the bursts we fitted the persistent spectrum just before each burst with a blackbody 
and a power law. During the burst, we fixed the power-law index and normalization to the value of
the persistent emission, and let the blackbody parameters to vary. Using this procedure we 
found results that are consistent with the ones we found using the standard method
therefore, in the rest of the paper, we used the results we obtained using the standard method

We defined the cooling phase for  different types of bursts observed in 4U 1636-53
as follows: The cooling phase of non-PRE single-peaked and PRE bursts starts from the 
moment that the fitted blackbody temperature starts to decrease until the end
of the burst, defined as the time when the burst flux is within
3 $\sigma$ of the pre-burst persistent emission. For double-peaked bursts we 
used two different definitions of the cooling phase,  one that starts  
from the moment the blackbody temperature starts to decrease after the first peak 
of the burst until the end of the burst, and the other one from the moment the 
blackbody temperature starts to decrease after the second peak of the burst until 
the end of the burst. We did not consider the triple-peaked burst reported in
\cite{zhang gb}

\section{result}
\label{results}

\subsection{Temperature distribution during the cooling phase of X-ray bursts}

\begin{figure}
\centering
\includegraphics[width=60mm,angle=270]{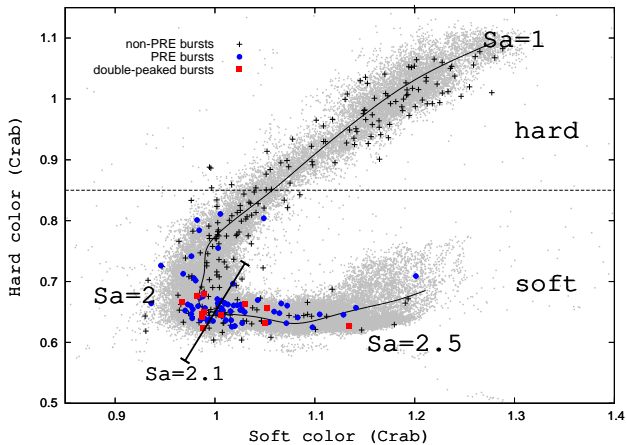}
\caption{ Colour-colour diagram of all RXTE observations of 4U 1636-53 up to May 2010. The grey points represent the colours of the source from all available RXTE observations. Each point in this diagram represents 256 s of data. The black crosses represent the colours of the persistent emission of the source at the onset of a single-peaked non-PRE X-ray burst. The filled blue circles indicate the same for PRE burst, and the red squares for double-peaked bursts. 
The position of 
the source on the diagram is parameterized by the length of the black solid curve $S_{\rm a}$. 
The horizontal line divides hard and soft state of the source (see text for details). The diagonal solid bar
divides the PRE bursts into
two groups ($S_{\rm a}>2.1$ and $S_{\rm a}<2.1$; see text for details.) } 
\label{fig: CCD}
\end{figure}

\begin{figure}
\centering
\includegraphics[width=60mm,angle=270]{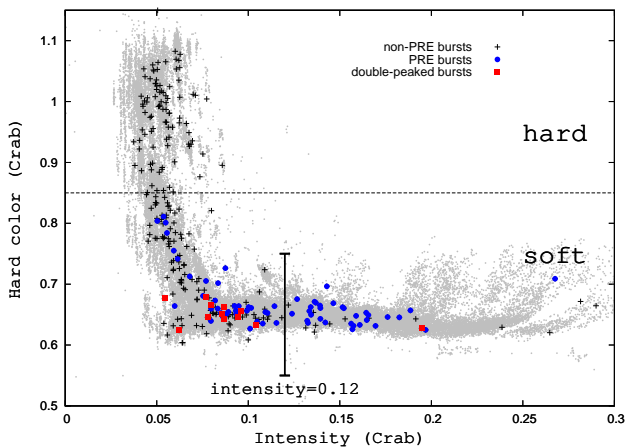}
\caption{ Hardness-intensity diagram of all RXTE observations of 4U 1636-53. The symbols are
the same as in Figure \ref{fig: CCD}. The PRE bursts are divided into two groups (intensity 
$> 0.12$ Crab and intensity $< 0.12$ Crab) by the vertical solid bar.} 
\label{fig: HID}
\end{figure}

In order to understand the cooling tails of PRE, single-peaked no-PRE,
and double-peaked bursts, we fitted the time-resolved spectra with an absorbed
blackbody and compared the 
fitting parameters. From the CD in Figure \ref{fig: CCD} and
hardness-intensity diagram (HID) in Figure \ref{fig: HID}, we found that all 
the PRE bursts and double-peaked bursts have a hard colour smaller than 0.85 
\citep[see also ][]{Muno04, zhang gb}. 
We therefore divided the bursts in four groups based on their persistent
hard colour and their time-resolved spectrum (see Figures \ref{fig: CCD} and
\ref{fig: HID}): (i) Hard non-PRE bursts; these are  single-peaked non-PRE bursts
that took place when the persistent hard colour of the source was larger 
than 0.85. (ii) Soft non-PRE bursts; these are single-peaked non-PRE bursts 
that took place when the persistent hard colour of the source was smaller than 
0.85. (iii), PRE bursts; (iv) Double-peaked bursts.

\begin{figure*}
    \centering
    \subfigure[non-PRE single-peaked bursts]
    {
        \includegraphics[width=2.10in,angle=270]{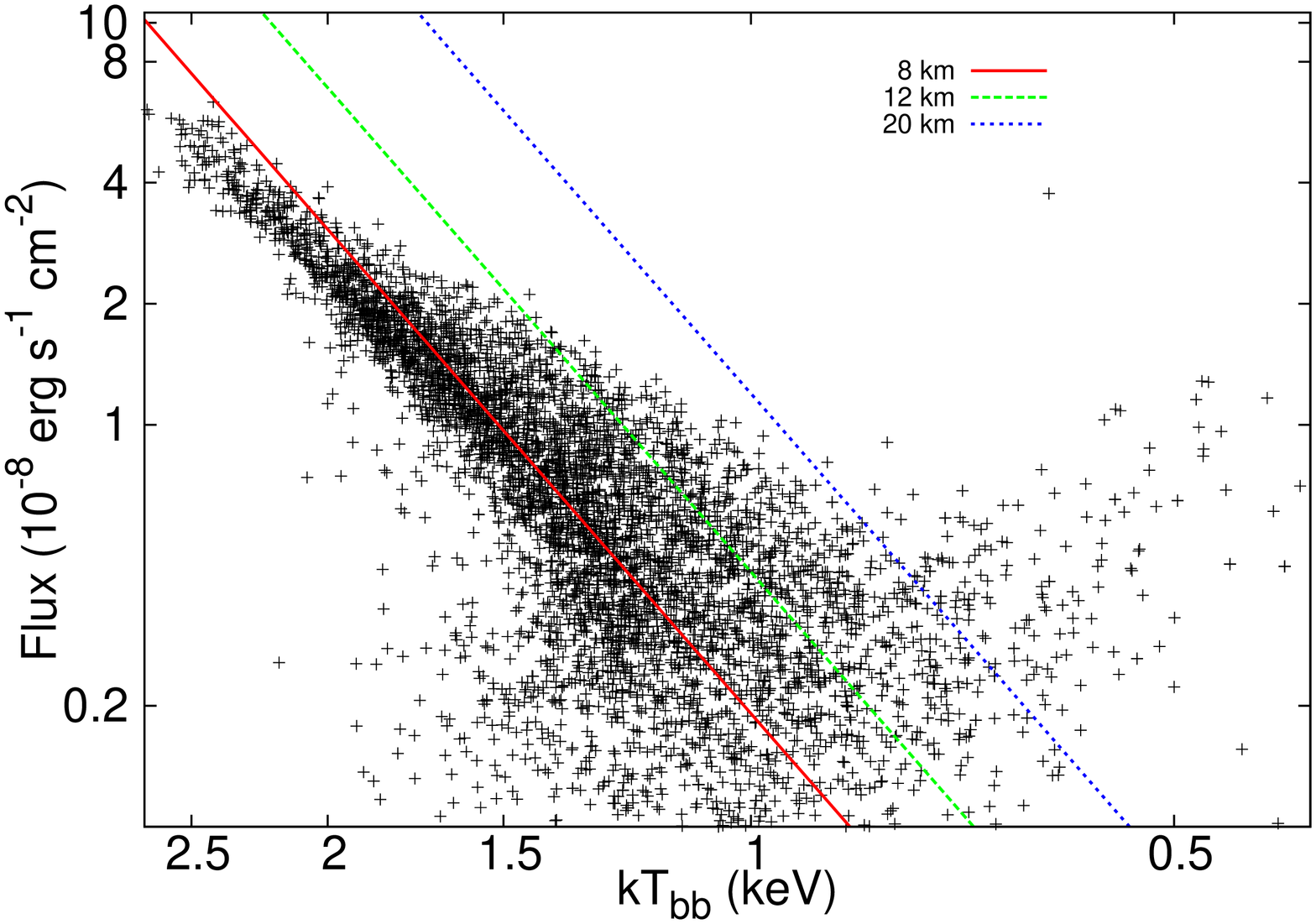}
        \label{fig:1_sub}
    }
    \subfigure[PRE bursts]
    {
        \includegraphics[width=2.1in,angle=270]{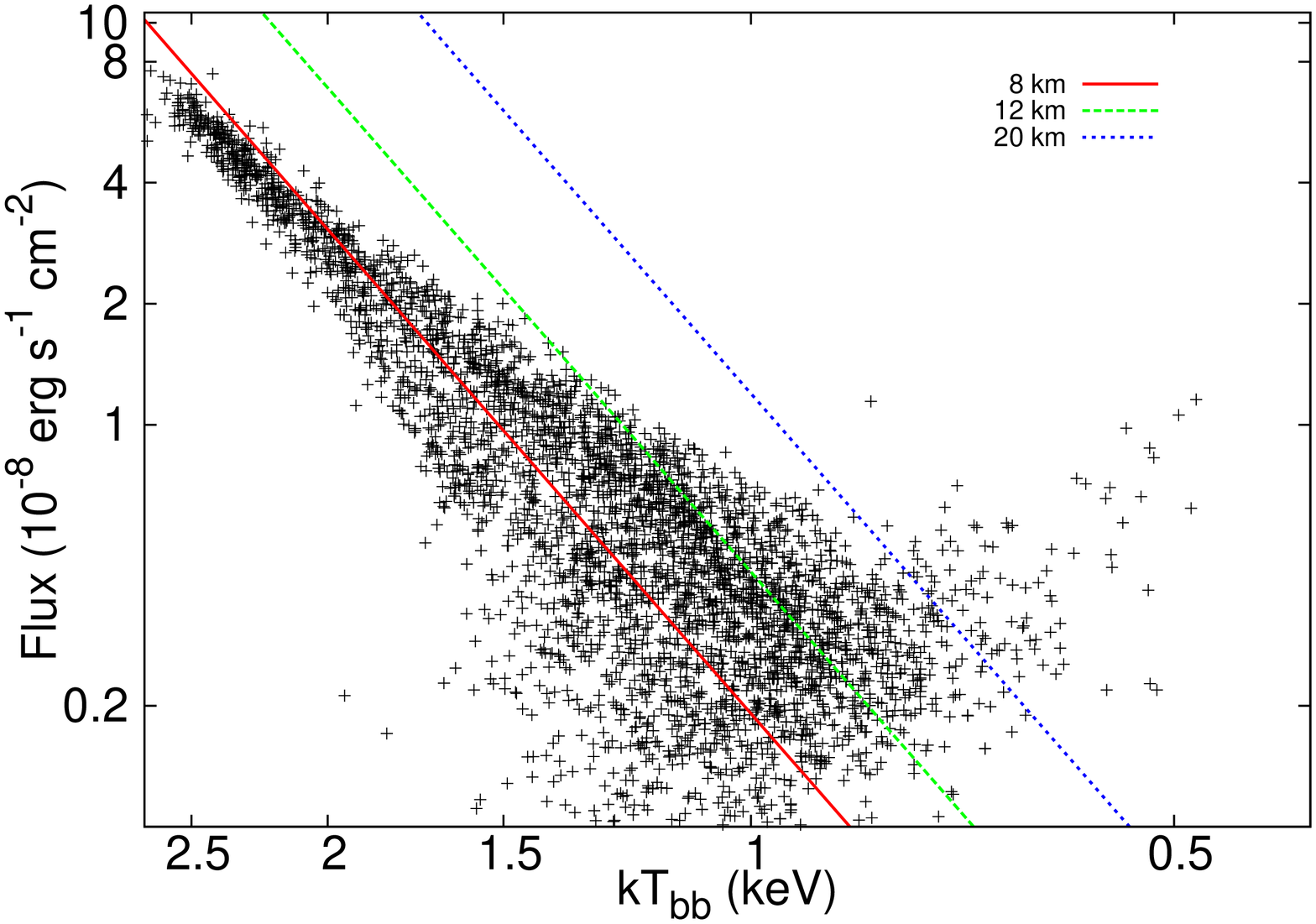}
        \label{fig:2_sub}
    }
    \subfigure[double-peaked bursts after first peak]
    {
        \includegraphics[width=2.1in,angle=270]{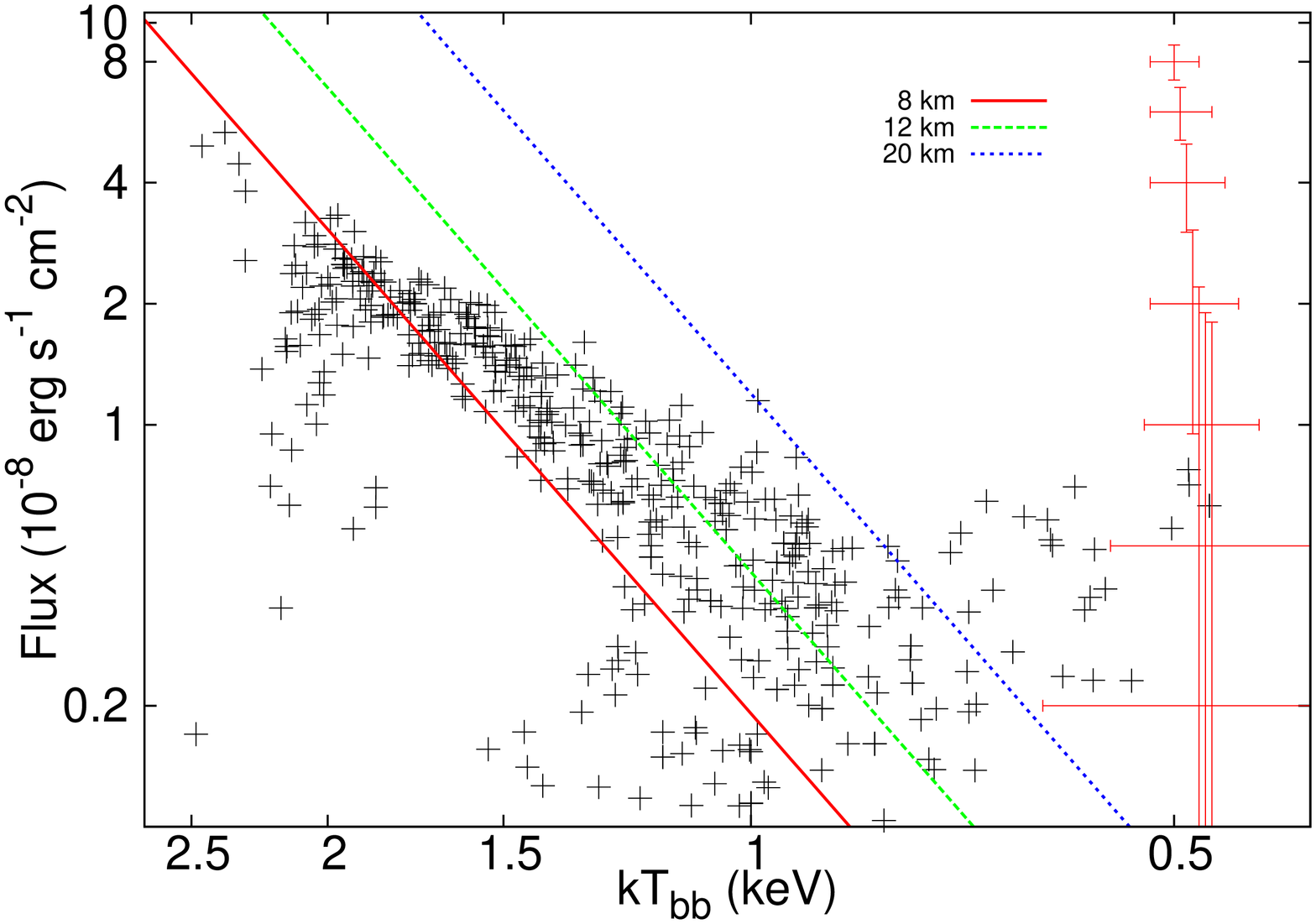}
        \label{fig:3_sub}
    }
    \subfigure[double-peaked bursts after second peak]
    {
        \includegraphics[width=2.1in,angle=270]{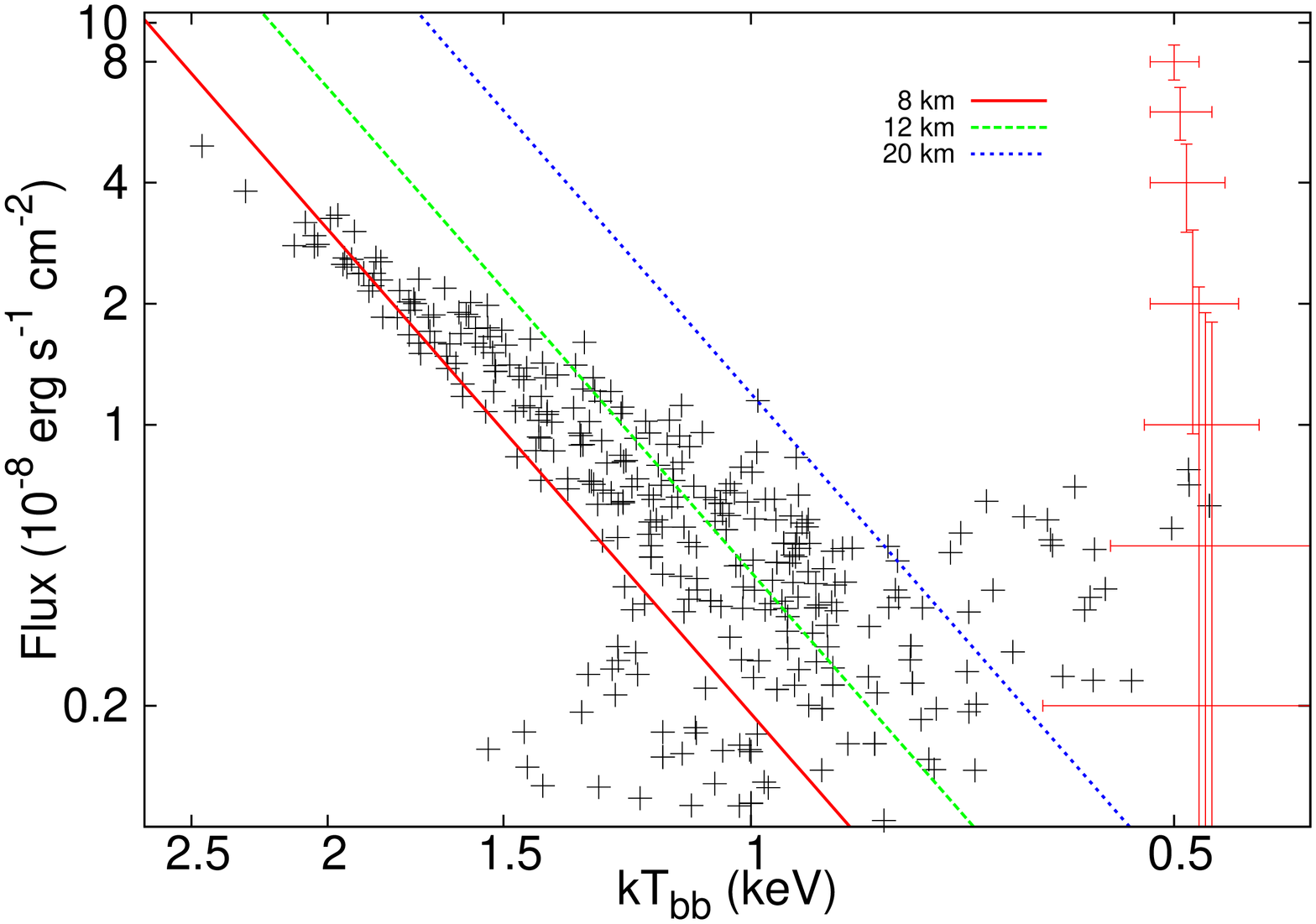}
        \label{fig:4_sub}
    }
    \caption{Bolometric flux as a function of blackbody colour temperature for different 
types of bursts in 4U 1636-53 during the cooling phase of the burst. Each point on the 
plot represent 0.5 seconds. The diagonal lines represent lines of constant inferred radius.
Typical 90\% confidence level error bars are indicated in the lower panels..
             }
    \label{fig:``H-R'' diagram}
\end{figure*}

In Figure \ref{fig:``H-R'' diagram}, we investigate the relation between 
the bolometric flux  and the color temperature during the cooling phase of the bursts 
using 0.5 s bins (see section 2). In this so-called flux-temperature (FT) diagram 
we also show lines of constant inferred radius, calculated from
$F/(\sigma T^{4})$. The top panels (a) and (b) of 
Figure \ref{fig:``H-R'' diagram} show the cooling phase of 
non-PRE bursts and PRE bursts, respectively; the bottom panels (c) and (d) show the
cooling phase of double-peaked bursts after the first and second peak, 
respectively. In order to compare different types of bursts at the same source state,
in panel (a) of Figure \ref{fig:``H-R'' diagram} we only plot the soft non-PRE bursts.
At the beginning of the cooling phase (upper left part of the panels), 
both PRE and soft non-PRE bursts appear to have 
similar colour temperature. However, at the end of the cooling phase (lower-right 
part of the panels) the soft non-PRE bursts
appear to have a larger spread of temperatures than the PRE bursts.
Comparing the two bottom panels in Figure \ref{fig:``H-R'' diagram}, we notice that
there are some points at high flux level but below the 8-km radius 
red line in panel (c) that are not present in panel (d). These points are the ones 
between the two peaks in double-peaked bursts. After the second peak, in double-peaked 
bursts most of the points appear above the red line.

\begin{table*}
\begin{center}
\caption{The average  and standard deviation of the temperature distribution for 
different types of bursts at different flux levels. }
\begin{tabular}{cccccccccccc}

\hline \hline
Flux range ($10^{-8}$ erg cm$^{-2}$ s$^{-1}$) & & 0-1.0   & 1.0-2.0  & 2.0-4.0  & 4.0-7.0   \\
\hline
hard non-PRE & $<kT_{\rm bb}>$  (keV)  & 1.42    & 1.79     & 2.05     & 2.37     \\
             & $\sigma$ (keV) & 0.27    & 0.15     & 0.18     & 0.64     \\
\hline
soft non-PRE & $<kT_{\rm bb}>$  (keV)  & 1.21    & 1.61     & 2.01     & 2.42     \\
             & $\sigma$ (keV) & 0.23    & 0.21     & 0.20     & 0.38     \\
\hline
PRE          & $<kT_{\rm bb}>$  (keV)  & 1.09    & 1.58     & 1.98     & 2.36     \\
             & $\sigma$ (keV) & 0.23    & 0.20     & 0.19     & 0.19     \\
\hline
double-peaked & $<kT_{\rm bb}>$  (keV)  & 0.98    & 1.46     & 1.83     & 2.45  \\
              & $\sigma$ (keV) & 0.23    & 0.23     & 0.38     & 1.20  \\

\hline 
\end{tabular}
\label{tab:kt-statistic}
\begin{tablenotes}
\item[]
\end{tablenotes}
\end{center}
\end{table*}

\begin{figure*}
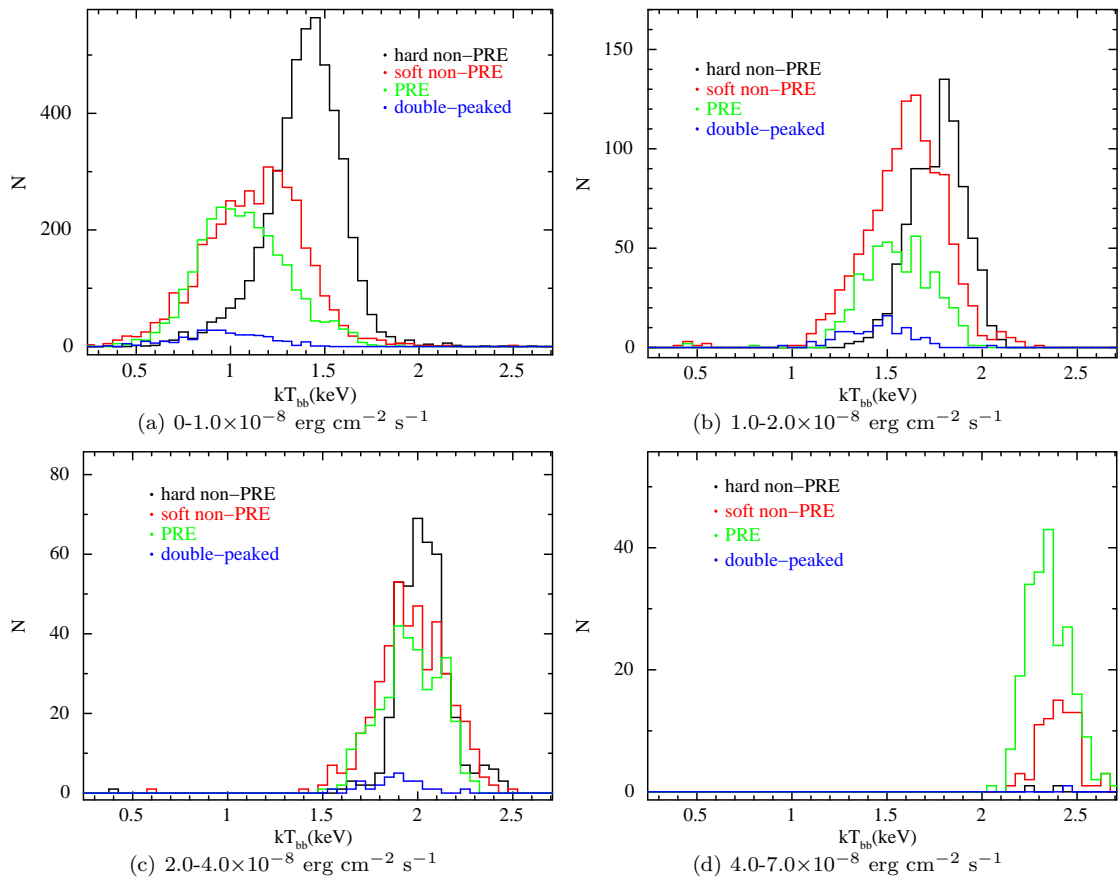

    \centering
    \subfigure[0-1.0$\times 10^{-8}$ erg cm$^{-2}$ s$^{-1}$]
    {
        \includegraphics[width=2.10in,angle=270]{qdp_npre_f0-1.ps}
        \label{fig:first_sub}
    }
    \subfigure[1.0-2.0$\times 10^{-8}$ erg cm$^{-2}$ s$^{-1}$]
    {
        \includegraphics[width=2.10in,angle=270]{qdp_npre_f1-2.ps}
        \label{fig:second_sub}
    }
    \subfigure[2.0-4.0$\times 10^{-8}$ erg cm$^{-2}$ s$^{-1}$]
    {
        \includegraphics[width=2.1in,angle=270]{qdp_npre_f2-4.ps}
        \label{fig:third_sub}
    }
    \subfigure[4.0-7.0$\times 10^{-8}$ erg cm$^{-2}$ s$^{-1}$]
    {
        \includegraphics[width=2.1in,angle=270]{qdp_npre_f4-7.ps}
        \label{fig: fourth_sub}
    }
    \caption{The distributions of colour temperatures during the cooling phase 
for hard non-PRE bursts (blue), soft non-PRE bursts (red), PRE bursts (black) and double-peaked bursts after the second peak (green), respectively. Each panel is for a different flux range,
as indicated.}
    \label{fig: histogram}
\end{figure*}

In order to study the distribution of temperatures in the cooling phase of the bursts
quantitatively, we divided the data of Figure
\ref{fig:``H-R'' diagram} in four different groups based on flux levels. 
Panels (a) to (d) in Figure \ref{fig: histogram} show the distribution of the colour 
temperature during the cooling phase of hard non-PRE bursts (black), soft non-PRE 
bursts (red), PRE bursts (green) and double-peaked bursts after the second peak (blue),
at different flux levels. In Table \ref{tab:kt-statistic} we give the 
average and standard deviation of the distribution of temperatures for each flux level.
We find that, as the bolometric flux decreases in the cooling tail, the average 
temperature of the hard non-PRE bursts becomes significantly higher than that of the soft
non-PRE bursts.  We also find that, as the flux decreases, the average temperature of 
soft no-PRE bursts becomes significantly higher than that of PRE bursts and double-peaked
bursts. At higher flux levels, the PRE bursts dominate the sample.

\begin{table*}
\begin{center}
\caption{KS test probabilities for the distributions in Figure \ref{fig: histogram}}
\begin{tabular}{ccccccccccc}

\hline \hline
Flux range ($10^{-8}$ erg cm$^{-2}$ s$^{-1}$)& 0$-$1.0            & 1.0$-$2.0            & 2.0$-$4.0            & 4.0$-$7.0   \\
\hline
soft non-PRE vs hard non-PRE  &  $0$  & $0$ & $3.0\times 10^{-8}$  &  $7.3\times 10^{-1}$   \\
\hline
soft non-PRE vs PRE & $0$  & $2.1 \times 10^{-6}$ & $1.5\times 10^{-1}$  &  $0.4\times 10^{-1}$   \\
\hline
hard non-PRE vs PRE & $0$  & $0$   & $6.2\times 10^{-9}$  &  $9.8\times 10^{-1}$   \\
\hline
soft non-PRE vs double-peaked & $0$ & $4.7\times 10^{-12}$ & $2.9\times 10^{-2}$  &  $6.1\times 10^{-1}$   \\
\hline
PRE vs double-peaked  & $1.1\times 10^{-7}$  & $3.2\times 10^{-5}$  & $3.9\times 10^{-2}$  &  $3.5\times 10^{-1}$   \\
\hline 
\end{tabular}
\label{tab:ks-test_1}
\begin{tablenotes}
\item[]
\end{tablenotes}
\end{center}
\end{table*}

\begin{table*}
\begin{center}
\caption{KS test probabilities after we have aligned  all the distributions to the same average 
temperature. }
\begin{tabular}{ccccccccccc}

\hline \hline
Flux range($10^{-8}$ erg cm$^{-2}$ s$^{-1}$)& 0$-$1.0  & 1.0$-$2.0            & 2.0$-$4.0            & 4.0$-$7.0   \\
\hline
soft non-PRE vs hard non-PRE  & $2.27\times10^{-7}$ & $3.0 \times 10^{-5}$ & $1.1\times 10^{-2}$  &  $9.3\times 10^{-1}$   \\
\hline
soft non-PRE vs PRE    & $4.6\times10^{-5}$  & $1.1 \times 10^{-2}$ & $1.8\times 10^{-1}$  &  $8.4\times 10^{-1}$   \\
\hline
hard non-PRE vs PRE    & $5.8\times10^{-11}$ & $1.6\times 10^{-2}$  & $1.9\times 10^{-3}$  &  $9.5\times 10^{-1}$   \\
\hline
soft non-PRE vs double-peaked & $7.8\times 10^{-1}$ & $5.9\times 10^{-1}$  & $1.3\times 10^{-2}$  &  $7.9\times 10^{-1}$   \\
\hline
PRE vs double-peaked  & $1.8\times 10^{-2}$ & $7.8\times 10^{-1}$  & $4.5\times 10^{-2}$  &  $8.3\times 10^{-1}$   \\
\hline 
\end{tabular}
\label{tab:ks-test_2}
\begin{tablenotes}
\item[]
\end{tablenotes}
\end{center}
\end{table*}

We carried out a Kolmogorov--Smirnov (KS) test to assess whether any two distributions
 are consistent with being the same. The results of
the KS tests are shown in Table \ref{tab:ks-test_1}. We find that, except at the highest
flux level, it is unlikely that any two sample distributions in Figure
\ref{fig: histogram} come from the same parent population. We note that there are 
only eleven double-peaked bursts in our sample and therefore the KS-test results are less
reliable when we compare other samples to this type of bursts. Since it is apparent 
that the average temperature of different types of bursts is 
different (Table \ref{tab:kt-statistic}), we repeated the KS test after we first 
aligned all distributions such that they all have the same average temperature. 
The new KS-test results are  shown in Table \ref{tab:ks-test_2}.  As expected, after
aligning the distributions, the probabilities of the KS test increased. But at low flux 
level, the probability that two samples come from the same parent population remains
small. This indicates that in these cases, not only the average,  but also the shape
of the distributions is different.

\subsection{Flux-Temperature relation during the cooling phase of X-ray bursts}
\label{sub f_t}
The evolution of the bolometric flux and the blackbody temperature during 
the cooling tails of all X-ray bursts in our sample appear to follow a power-law relation. 
We fitted a power law, 
$F_{\rm b} = \alpha T^{\gamma}_{\rm bb}$,  to the data for PRE, soft non-PRE and 
hard non-PRE bursts, separately, where $F_{\rm b}$ is the bolometric
flux, $T_{\rm bb}$ is the blackbody colour temperature, $\alpha$ is the normalization and
$\gamma$ is the power-law index. 

\begin{figure*}
    \centering
    \subfigure[all flux]
    {
        \includegraphics[width=1.40in,angle=270]{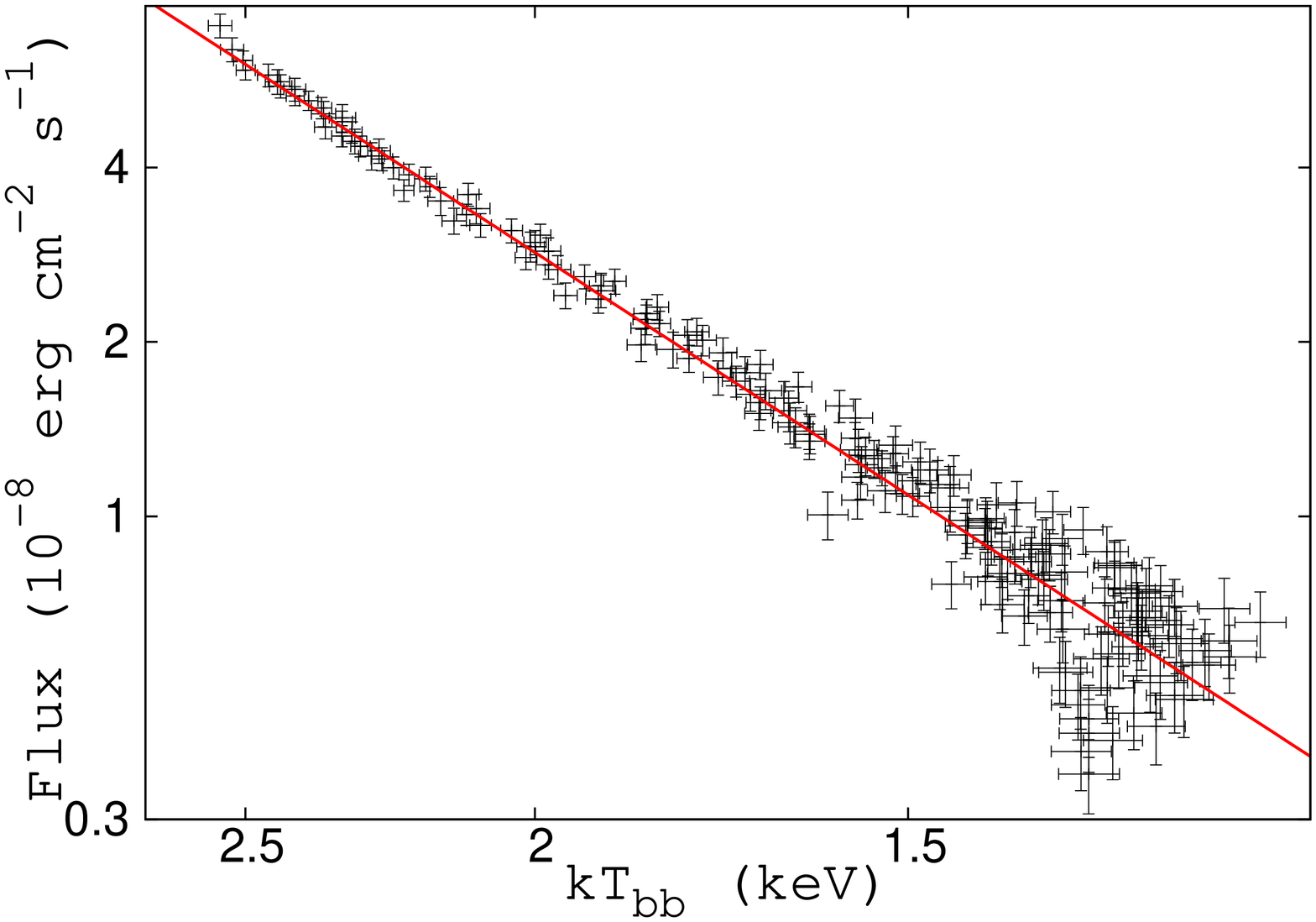}
        \label{fig:7_sub}
    }
    \subfigure[flux $>$ 1]
    {
        \includegraphics[width=1.40in,angle=270]{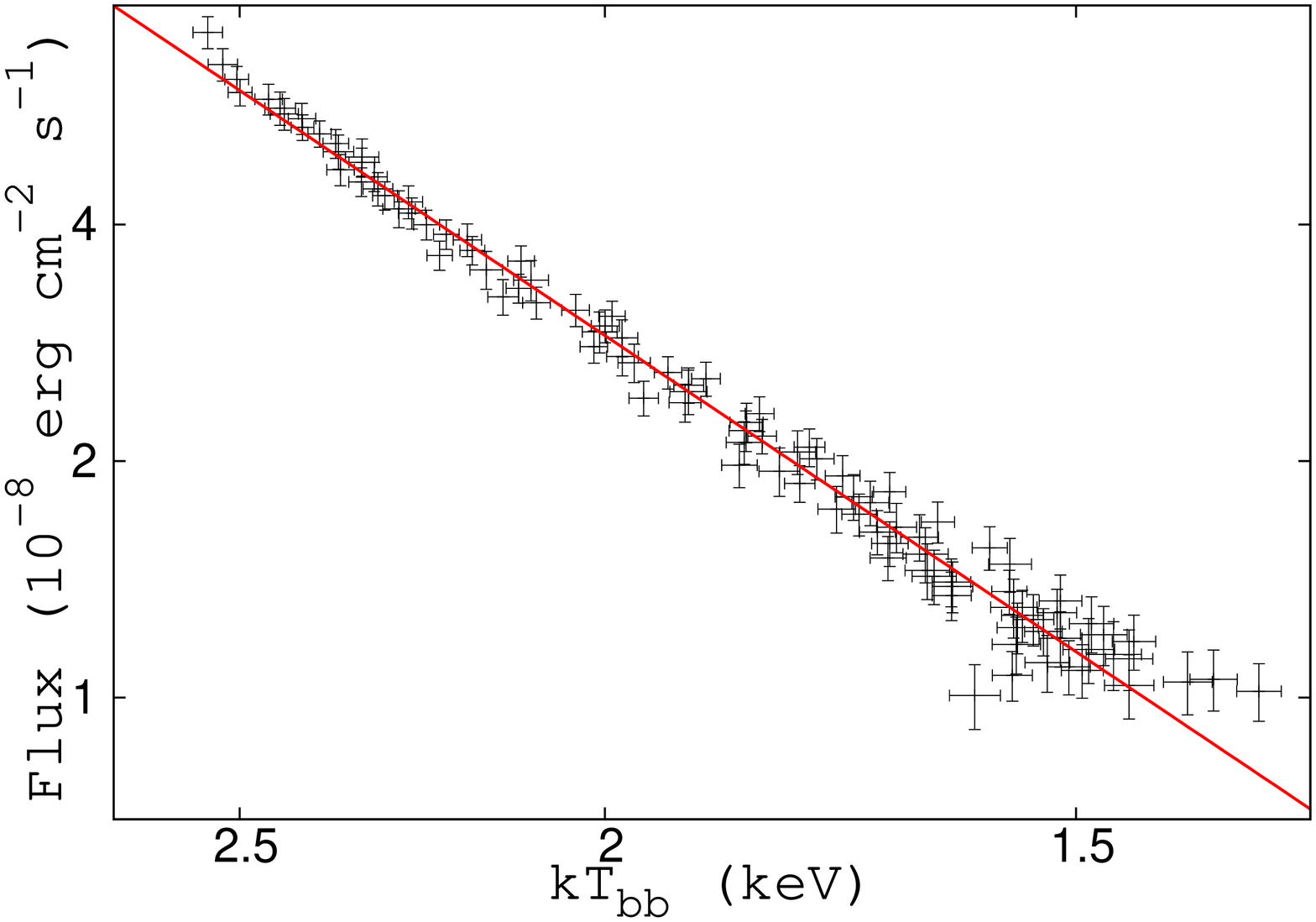}
        \label{fig:8_sub}
    }
    \subfigure[flux $>$ 2]
    {
        \includegraphics[width=1.40in,angle=270]{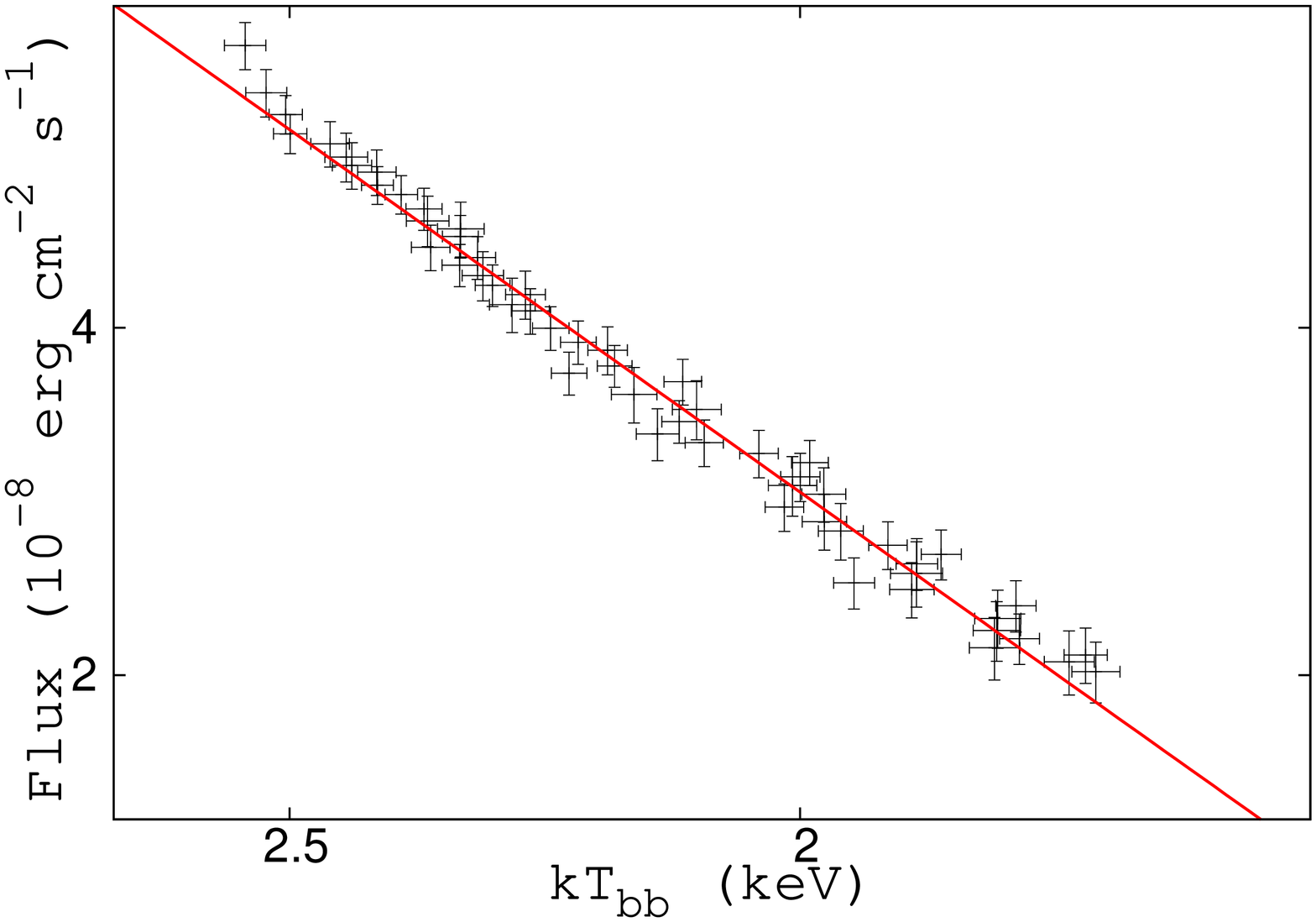}
        \label{fig:9_sub}
    }

    \subfigure[all flux]
    {
        \includegraphics[width=1.40in,angle=270]{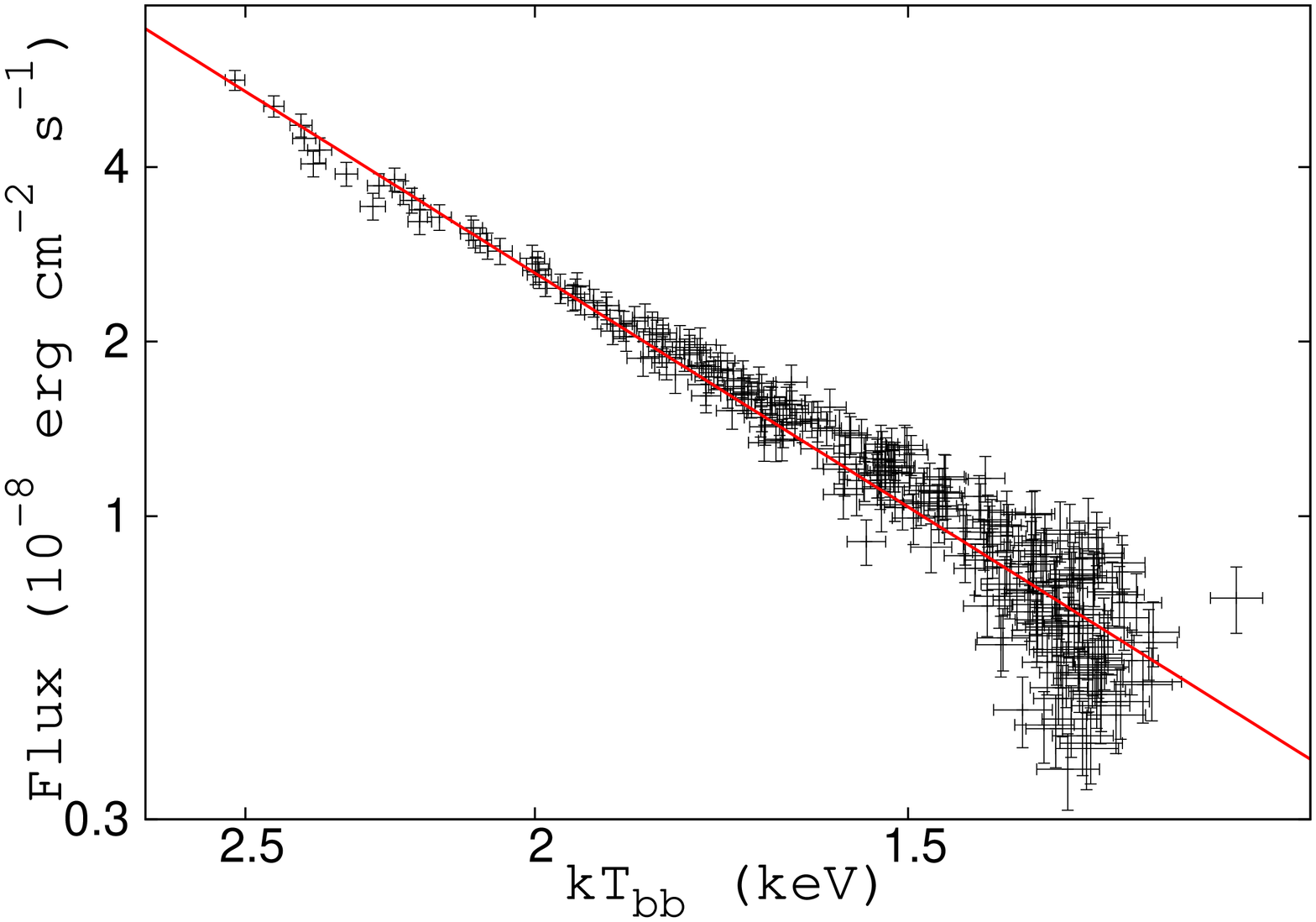}
        \label{fig:7_sub}
    }
    \subfigure[flux $>$ 1]
    {
        \includegraphics[width=1.40in,angle=270]{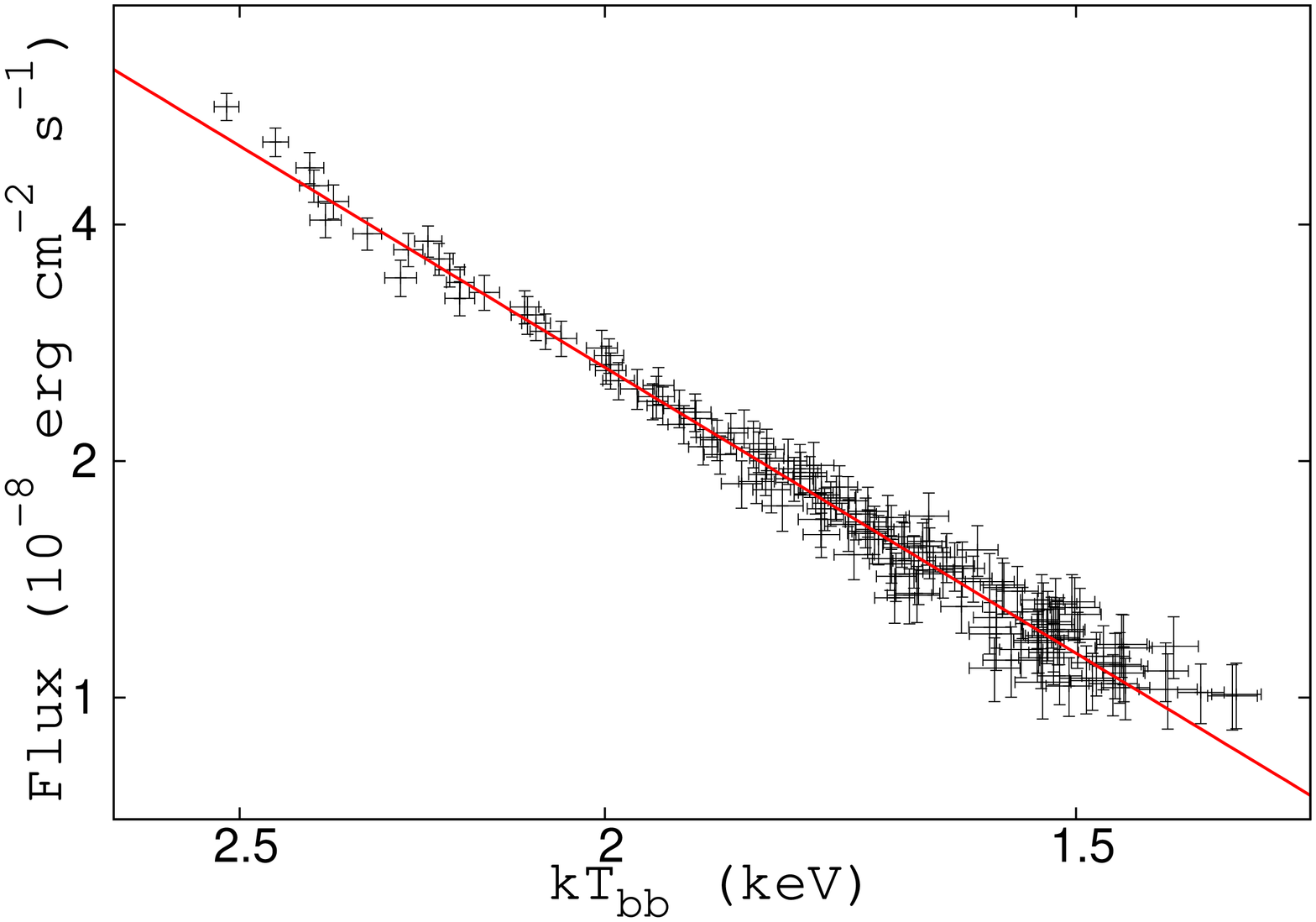}
        \label{fig:8_sub}
    }
    \subfigure[flux $>$ 2]
    {
        \includegraphics[width=1.40in,angle=270]{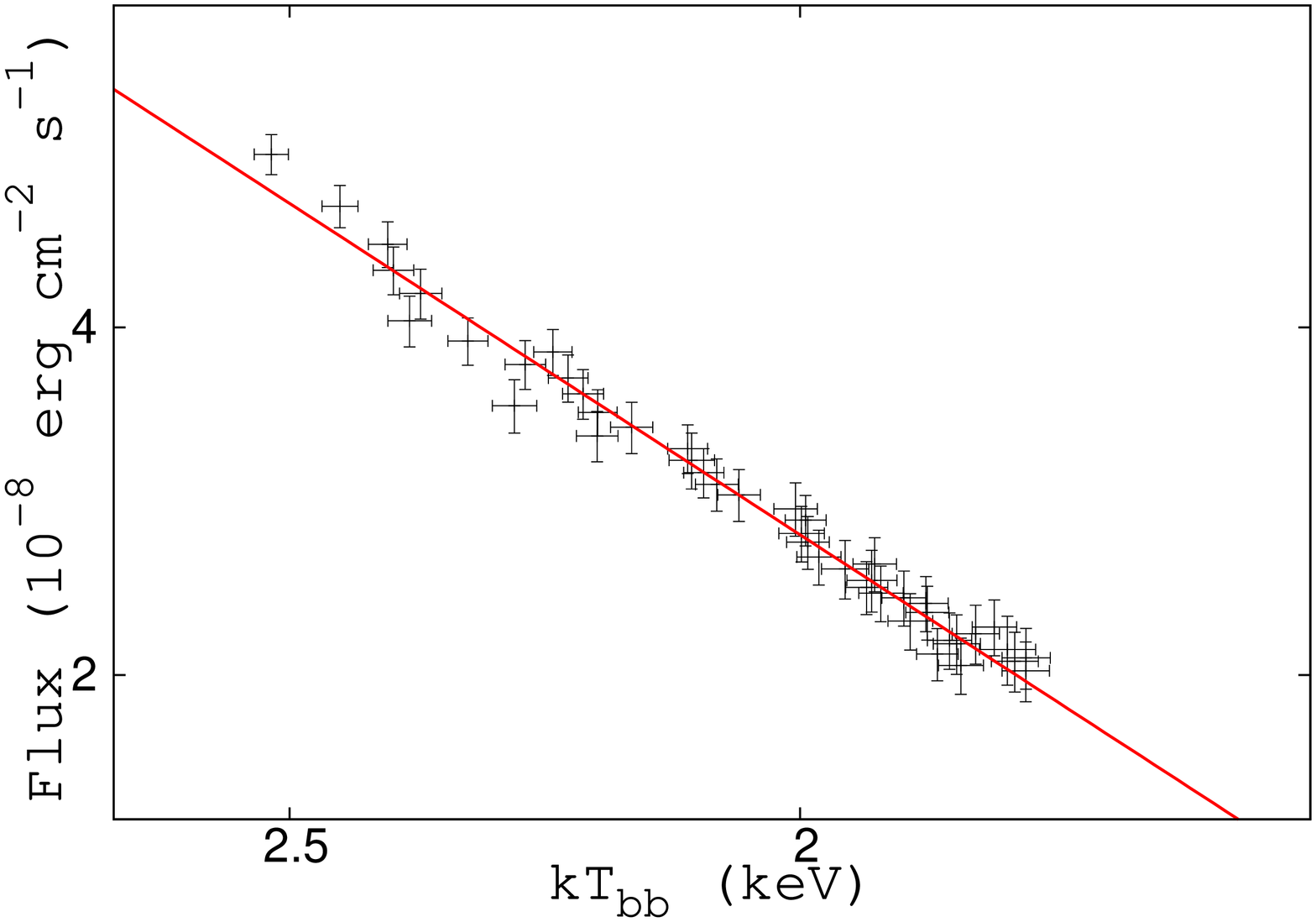}
        \label{fig:9_sub}
    }

    \subfigure[all flux]
    {
        \includegraphics[width=1.40in,angle=270]{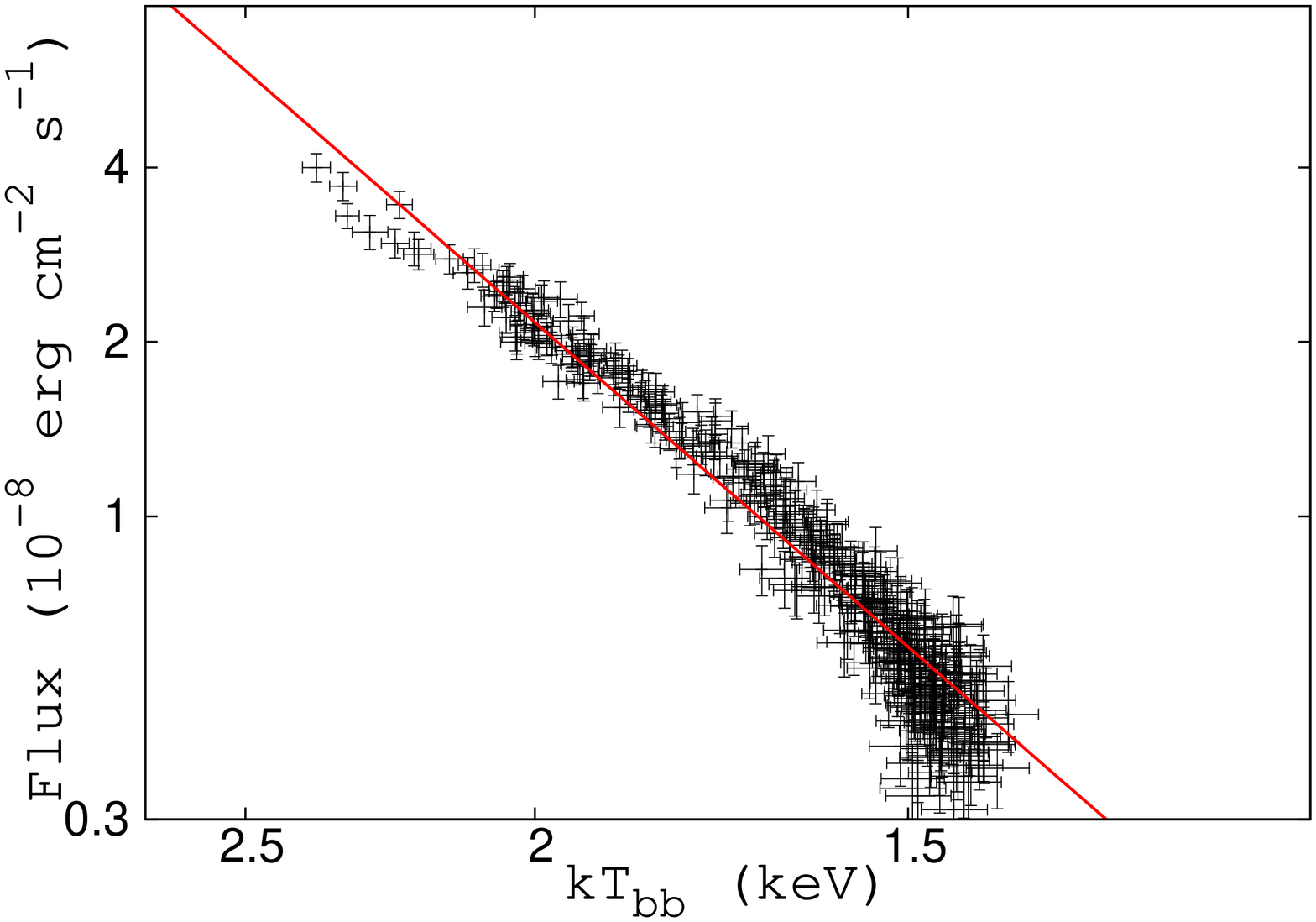}
        \label{fig:7_sub}
    }
    \subfigure[flux $>$ 1]
    {
        \includegraphics[width=1.40in,angle=270]{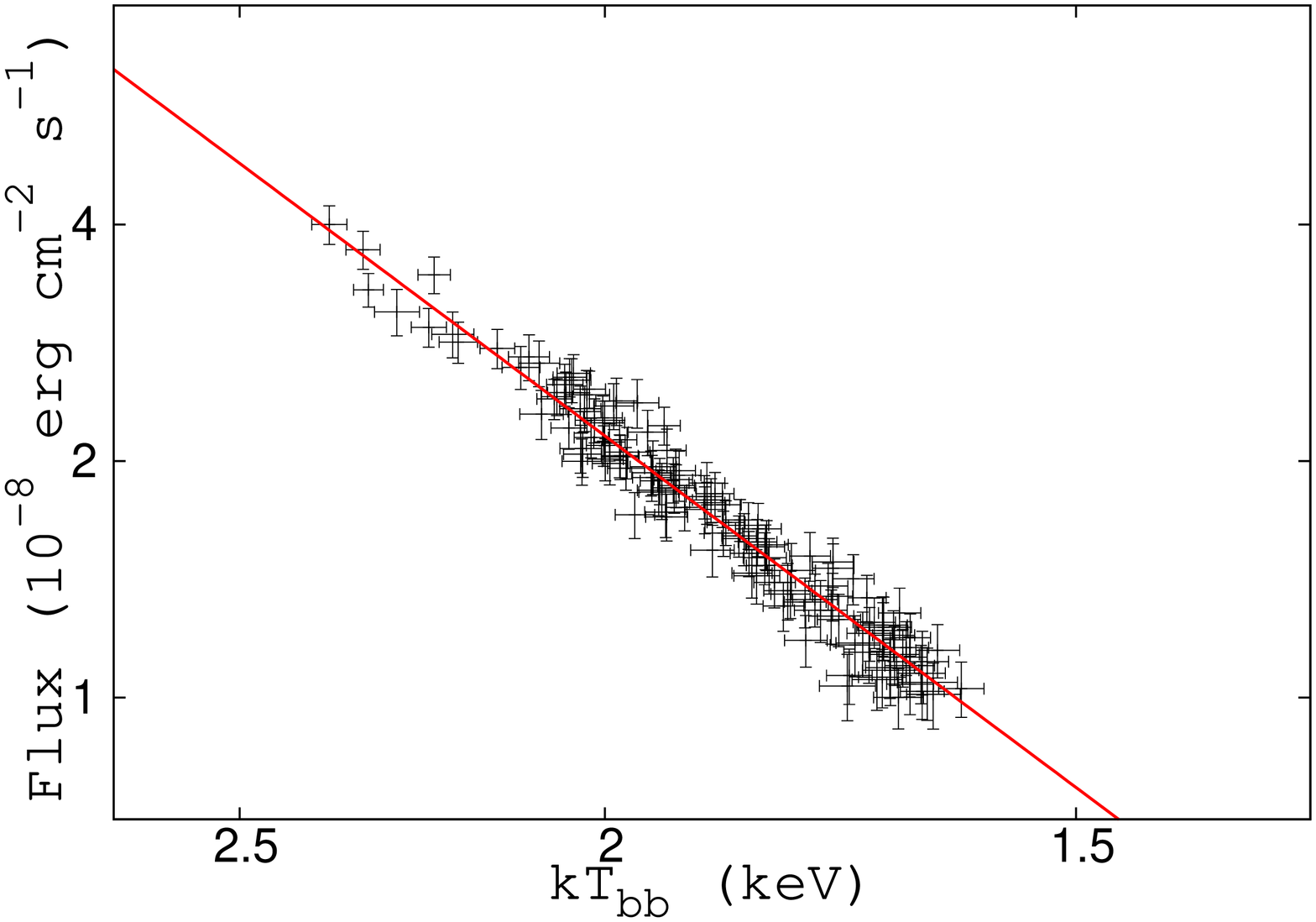}
        \label{fig:8_sub}
    }
    \subfigure[flux $>$ 2]
    {
        \includegraphics[width=1.40in,angle=270]{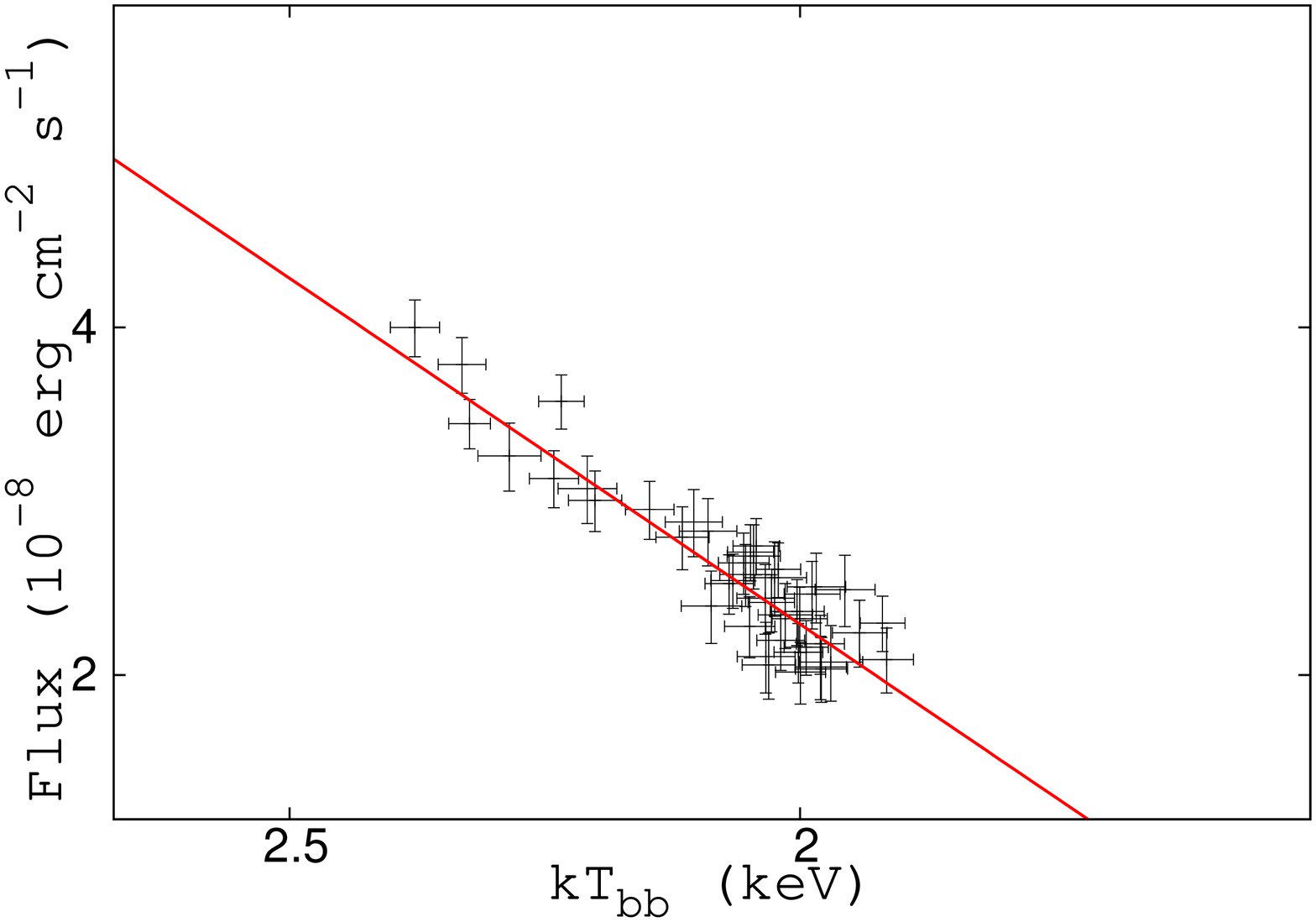}
        \label{fig:9_sub}
    }
    \caption{The flux-temperature diagram of PRE (top panels), soft non-PRE (middle panels) and 
hard non-PRE bursts (bottom panels). From left to right the panels show, respectively, all data, 
data where the flux was larger than $1\times 10^{-8}$ erg cm$^{-2}$ s$^{-1}$, and  data where
the flux was larger than $2\times 10^{-8}$ erg cm$^{-2}$ s$^{-1}$.
The best-fitting power law for each selection is shown with the red lines. 
}
    \label{fig: ft_fit}
\end{figure*}

For all type of burst, we first selected  data with $\frac{ F_{\rm b} }{ \delta F_{\rm b} } > 2$, 
where $\delta F_{\rm b}$ is the error of the bolometric flux, and we rebined the data by a  factor 10 
(we also tried $\frac{ F_{\rm b} }{ \delta F_{\rm b} } > 1$ and $\frac{ F_{\rm b} }{ \delta F_{\rm b} } > 3$,
but the power-law fitting results are consistent with the ones we found using 
$\frac{ F_{\rm b} }{ \delta F_{\rm b} } > 2$ ). 
We then divided the data into three groups: data with 
$F_{\rm b} > 1\times10^{-8}$ erg cm$^{-2}$ s$^{-1}$, 
$F_{\rm b} > 2\times10^{-8}$ erg cm$^{-2}$ s$^{-1}$, and all data, and  fitted each group 
separately. The best fitting results are shown in Table \ref{tab: cooling_fit}, 
and the fits are shown in Figure \ref{fig: ft_fit}.

A power law fits well the data with flux larger than $1\times10^{-8}$ erg cm$^{-2}$ s$^{-1}$
for all type of bursts, but a power law does not fit all the data (without flux selection) for
any of the three types of bursts (see Table \ref{tab: cooling_fit}). We also found that: (i)
During the cooling phase, the X-ray bursts of 4U 1636--53 do not follow the expected 
$F_{\rm b} \propto T^{4}_{\rm bb}$ relation; (ii) within the same type of bursts, the power-law index changes
significantly as the flux increases; (iii) within the same flux selection, the power-law 
index of hard non-PRE bursts is significantly higher than that of PRE and soft non-PRE 
bursts (except at the highest flux level where the fitting errors are large due to the 
limited data). Given that $L_{\rm b}=\alpha T^{\gamma}_{\rm bb}$, this means that for a 
fixed flux level, hard non-PRE bursts have higher 
$T_{\rm bb}$ than PRE and soft non-PRE bursts as the neutron star cools down or, for
a fixed $T_{\rm bb}$ during the cooling phase, the apparent emitting area of PRE and soft non-PRE
bursts is larger than that of hard non-PRE bursts. 


\begin{table*}
\begin{center}
\tiny
\caption{Power-law fit for different types of bursts in the flux-temperature diagram. In the table, 
$\alpha$ and $\gamma$ are the normalization and the index of the best-fitting power law. }
\begin{tabular*}{1.\textwidth}{c | c c c | c c c | c c c}
\hline \hline
Flux range                              &   & all  &    &  & $>$ 1.0  &    &   & $>$ 2.0 & \\
$10^{-8}$ erg cm$^{-2}$ s$^{-1}$        &      &   &    &         &   &    &          & & \\

\hline
             & $\alpha$ & $\gamma$ & $\chi^{2}/d.o.f$ &  $\alpha$      & $\gamma$ &  $\chi^{2}/d.o.f$ & $\alpha$    & $\gamma$ & $\chi^{2}/d.o.f$  \\
\hline 
PRE         & 0.28$\pm$0.01 & 3.35$\pm$0.03 & 420/182 & 0.31$\pm$0.01 & 3.22$\pm$0.06 
& 120/103 & 0.30$\pm$0.03 & 3.24$\pm$0.11 & 42/55 \\
soft non-PRE   & 0.28$\pm$0.01 & 3.23$\pm$0.04 & 627/256 & 0.35$\pm$0.02 & 2.91$\pm$0.06 
& 117/147 & 0.34$\pm$0.03 & 2.96$\pm$0.13    & 26/44\\
hard non-PRE   & 0.097$\pm$0.004 & 4.48$\pm$0.06 & 719/321 & 0.18$\pm$0.01 & 3.58$\pm$0.12 
& 139/133 & 0.26$\pm$0.06 & 3.09$\pm$0.31 & 49/44 \\
\hline
\end{tabular*}
\label{tab: cooling_fit}
\begin{tablenotes}
\item[]
\end{tablenotes}
\end{center}
\end{table*}

\subsection{Photospheric radius expansion bursts}

\begin{table*}
\begin{center}
\caption{KS test probabilities for low $S_{\rm a}$ ($<2.1$) and high $S_{\rm a}$
($>2.1$) PRE bursts, and bright ($I > 0.12$ Crab) and faint ($I < 0.12$ Crab) PRE bursts. }
\begin{tabular}{ccccccccccc}
\hline \hline
Flux range ($10^{-8}$ erg cm$^{-2}$ s$^{-1}$) & 0$-$1.0 & 1.0$-$2.0   & 2.0$-$4.0 \index{\footnote{}} & 4.0$-$7.0   \\
\hline
low $S_{\rm a}$ vs high $S_{\rm a}$ & $7.6\times 10^{-14}$ & $5.0 \times 10^{-2}$ & $1.3\times 10^{-2}$  &  $3.6\times 10^{-1}$   \\
\hline
bright  vs faint  & $9.1\times 10^{-15}$ & $3.6 \times 10^{-1}$ & $4.8\times 10^{-1}$  &  $1.3\times 10^{-2}$   \\
\hline
\end{tabular}
\label{tab:ks-test_3}
\begin{tablenotes}
\item[]
\end{tablenotes}
\end{center}
\end{table*}

\begin{figure*}
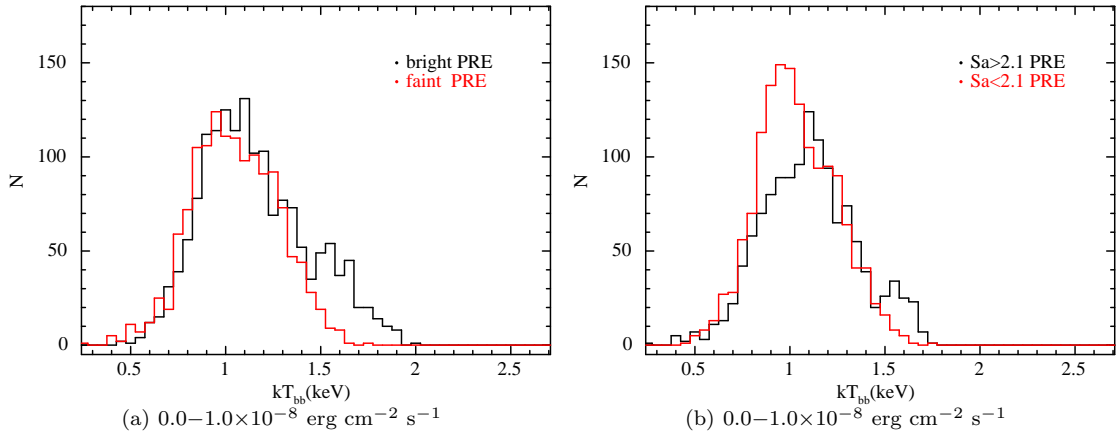

    \centering
    \subfigure[0.0$-$1.0$\times 10^{-8}$ erg cm$^{-2}$ s$^{-1}$]
    {
        \includegraphics[width=2.10in,angle=270]{qdp_pre_int_f0-1.ps}
        \label{fig:1_sub}
    }
    \subfigure[0.0$-$1.0$\times 10^{-8}$ erg cm$^{-2}$ s$^{-1}$]
    {
        \includegraphics[width=2.10in,angle=270]{qdp_pre_sz_f0-1.ps}
        \label{fig:2_sub}
    }
    \caption{The temperature distribution during the cooling phase of PRE bursts with 
Flux $< 1\times 10^{-8}$ erg cm$^{-2}$ s$^{-1}$. Panel (a) shows PRE bursts with 
persistent intensity $>$ 0.12 Crab (black) and $<$ 0.12 
Crab (red); panel (b) shows the PRE bursts with the $S_{\rm a}$ value
$>$ 2.1 (black) and $>$ 2.1 (red).}
    \label{fig: hist_int}
\end{figure*}

The PRE bursts concentrate on a narrow range of hard colours in the CD and HID,
but they span a large range of soft colours in the CD, and a large range of intensity 
in the HID. We therefore divided the PRE bursts in two groups according to the 
position of the source in the CD and HID at the time the bursts started, one group
with $S_{\rm a} < 2.1 $ (low-$S_{\rm a}$ PRE bursts) and the other with  
$S_{\rm a} > 2.1 $ (high-$S_{\rm a}$ PRE bursts; see the solid bar 
in Figure \ref{fig: CCD}). We also 
divided the PRE bursts in two groups according to the intensity of the persistent 
emission of the source at the time the burst started (see Figure \ref{fig: HID}): 
the faint PRE burst, with an intensity lower than 0.12 Crabs, and the
bright PRE burst, with an intensity higher than 0.12 Crabs. In panel (a) of Figure 
\ref{fig: hist_int} we show the distribution of colour temperature during the cooling
phase of bright PRE (black) and faint PRE bursts (red), 
and for low-$S_{\rm a}$ PRE (red) and high-$S_{\rm a}$ PRE bursts (black) in panel (b) of Figure 
\ref{fig: hist_int}; in both panel we selected only data with flux less than  
$1 \times 10^{-8}$ erg cm$^{-2}$ s$^{-1}$. The KS test for these two groups for all
flux levels, is shown in Table \ref{tab:ks-test_3}.  
The KS test still gives small probabilities at the lowest flux level; at other flux
levels the samples are consistent with coming from the same parent populations.

\section{Discussion}
\label{discussion}

During their cooling phase, the X-ray bursts in 4U 1636--53 do not
follow the bolometric flux-temperature relation $F_{\rm b} \propto T^4_{\rm bb}$, 
which would be expected if the
apparent emitting area on the neutron star remained constant during this
phase. The relation between bolometric flux and temperature is significantly
different between photospheric (PRE) and non-photospheric  (non-PRE)
bursts, as well as between non-PRE bursts that happen when the source is in
the hard state (hard non-PRE bursts) and the soft state (soft non-PRE
bursts). We also found that the temperature distribution at different flux
levels during the cooling phase is significantly different between PRE,
hard non-PRE and soft non-PRE bursts. This is consistent with the fact 
that different types of bursts have different power-law indices in the
flux-temperature diagram. 


This result \citep[see also][]{Gottwald, Bhattacharyya10, Suleimanov} is at
variance with the findings of \cite{Guver10}, who studied
three PRE bursts in  another source, 4U 1820--30, and found that these
PRE bursts follow the $F_{\rm b} \propto T^4_{\rm bb}$ relation quite well. \cite{Guver10}, 
however, did not consider two other  PRE bursts in
this source because the cooling phase of these two bursts showed a
complex behaviour, in which the emitting neutron-star area appeared not
to be constant. Here we studied 65 PRE bursts and 223
non-PRE bursts (out of which 94 and 129  were, respectively,
hard non-PRE and soft non-PRE bursts), and we did not discard any burst
in our sample. While some of the individual bursts in 4U 1636--53 could
lay close to the $F_{\rm b} \propto T^4_{\rm bb}$ relation (see Fig. \ref{fig:``H-R'' diagram}), 
several others deviate significantly from that relation, and on average the
relation is significantly different from $F_{\rm b} \propto T^4_{\rm bb}$ for any type of
burst in our sample. The discrepancy between our results and those of
\cite{Guver10} could be due to differences between the two sources: 4U
1820--30 is an ultracompact binary, with an orbital period of 685s, in 
which the neutron star accretes He from the companion \citep{Stella, Cumming}, 
whereas 4U 1636--53 has a longer orbital period (3.8 hours; see
\S 1), and the accreted material is mostly hydrogen. It is also possible
that the results of \cite{Guver10} are biased by the small
number of bursts they used, three PRE bursts in their case compared to 65 in
our case, or by the fact that they ignored two of the PRE bursts in 4U
1820--30 for their analysis.

Time-resolved spectra in the cooling phase of thermonuclear X-ray bursts
can be used to measure the radii and masses of neutron stars. The net
spectra of the thermonuclear X-ray bursts are usually well described by
a blackbody spectrum \citep{Strohmayer03, galloway}, which allows us to
obtain $R_{\rm bb}$, $T_{\rm bb}$, the blackbody radius and colour
temperature, respectively, from which we can calculate the bolometric
flux of the neutron star. The departure from the $F_{\rm b} \propto T^4_{\rm bb}$
relation in the cooling phase of all types of bursts in 4U 1636--53
means that the blackbody radius, $R_{\rm bb}$, is not constant during
the cooling phase of the bursts. This could be due to either changes in
the emitting area of the neutron star during this phase, or to changes in 
the colour-correction factor, which accounts for
hardening of the spectrum arising from electron scattering in the
neutron-star atmosphere  \citep{london86, Madej04} :  $f_{\rm c} = T_{\rm bb}/T_{\rm eff} = 
\sqrt{R_{\infty}/R_{\rm bb}}$, where $R_{\infty}$ is the neutron-star 
radius observed at infinity. 

If $f_{\rm c}$ remains constant throughout the cooling phase of all
X-ray bursts, the differences of $R_{\rm bb}$ should due to the changes
of the emission area on the neutron-star surface. 
From Figure \ref{fig:
ft_fit} and Table \ref{tab: cooling_fit}, it is apparent that the emission area in hard
non-PRE bursts is smaller than in PRE bursts and soft non-PRE bursts,
and that at the end of the cooling phase, the emitting area decreases
faster in hard non-PRE bursts than in the other types of bursts. 
\cite{Gottwald} also found that  
in EXO 0748-676 the average $R_{\rm bb}$ in the cooling phase increases as the  
persistent flux increases \cite[see also][]{Bildsten2000}. 
If for hard non-PRE bursts only a fraction of the surface of the 
neutron star is emitting during the cooling phase,
and the size of the emitting area decreases as the burst decays, one
would expect to see burst oscillations in the cooling tail of hard
non-PRE bursts. However, most bursts with oscillations in 4U 1636--53
occur when the source is in the soft state \citep{Muno04}. 

Alternatively, if $R_{\rm \infty}$ should remain constant throughout the
cooling phase of all the bursts, variations in $R_{\rm bb}$ can only be
due to changes of $f_{\rm c}$. We can write the colour-correction factor
as:

\begin{equation} \label{eq:fc1}
f_{\rm c}=\sqrt{\frac{R_{\rm \infty}}{d\sqrt{\frac{F}{\sigma T^{4}_{\rm
bb}}}}}=\sqrt{\frac{R(1+z)}{d\sqrt{\frac{F}{\sigma T^{4}_{\rm bb}}}}} ,
\end{equation}

where $d$ is the distance to the source, $R$ is the neutron-star radius,
$z$ is the gravitational redshift, and $\sigma$ is the Stefan-Boltzmann
constant. In Figure \ref{fig:
fc_model} we plot $f_{\rm c}$ as a function of the neutron-star
luminosity in Eddington units. For this plot we assumed $R=9$ km,
$(1+z)=1.35$, $d = 6.0$ Kpc, and we used  equation (\ref{eq:fc1}) and
the same data ($T_{\rm bb}$ and $R_{\rm bb}$) as in section \ref{sub f_t}. To
calculate the fluxes in Eddington units, $F_{\rm b}/F_{\rm Edd} = L_{\rm b}/L_{\rm
Edd}$, for PRE and soft non-PRE bursts we
used the average peak flux of PRE bursts as the Eddington flux, $F_{\rm
Edd}$. Assuming that PRE and soft non-PRE bursts are He bursts, and hard
non-PRE bursts are H bursts, for the hard non-PRE bursts we took 59\% of
the Eddington flux that we used in PRE bursts (assuming a hydrogen mass fraction,
 $X=0.7$).

\begin{figure}
\centering
\includegraphics[width=60mm,angle=270]{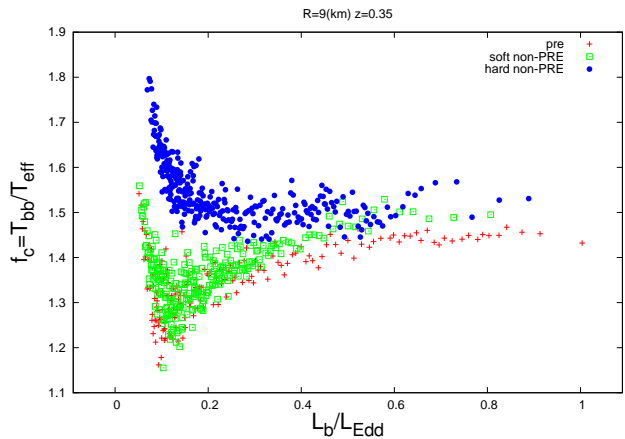}
\caption{Colour-correction factor as a function of luminosity in
Eddington units.  The different colour points correspond to different
types of bursts, as indicated in the plot. } 
\label{fig: fc_model}
\end{figure}

Figure \ref{fig: fc_model} shows that, as the luminosity increases, the
colour-correction factor of PRE and soft non-PRE bursts first decreases,
then increases, and finally remains more or less constant. For hard non-PRE 
bursts the colour-correction factor
first decreases, and then remains more or less constant as the
luminosity increases. At the same luminosity level, $f_{\rm c}$ in PRE
and soft non-PRE bursts is smaller than in hard non-PRE bursts, while
$f_{\rm c}$ is consistent with being the same in PRE and soft non-PRE
bursts. The data in Figure \ref{fig: fc_model} agree qualitatively with
the predictions of the model of \cite{Suleimanov} (see their Figure 2),
if we assume that during the cooling phase of hard non-PRE bursts the
neutron-star atmosphere is H rich, whereas for PRE and soft non-PRE
bursts the atmosphere is He rich. At high luminosity levels, the data
show that $f_{\rm c}$ stays constant, and does not increase as quickly
as predicted by the model. We speculate that during the early phases of
the cooling phase of all bursts, there is an enhanced abundance of heavy elements in
the neutron-star atmosphere that were produced during the thermonuclear
flash, and this effect is stronger in the PRE and soft non-PRE bursts where
dredge up of ashes is more effective \citep{Weinberg05}; as the neutron-star cools down, the heavy elements settle
down faster than H and He, and therefore the heavy-element abundance in
the neutron-star atmosphere decreases as the burst decays. This is reported by
the models of \cite{Suleimanov}, in which $f_{\rm c}$  decreases as the
high-element abundance in the neutron-star atmosphere increases. \cite[The
same is true for the model of][]{Guver10} It is then
possible that the difference at high luminosity between the data and the
models of \cite{Suleimanov} is due to changes of the chemical
composition of the neutron-star atmosphere during the early cooling
phase of the bursts. 

We also found a significant difference in the the distribution of the
blackbody temperature during the decay of X-ray bursts in 4U 1636--53 in
different types of bursts at different flux levels \citep[see also][]{int09, Bhattacharyya10}. 
At low flux level, the average $T_{\rm bb}$ is higher in hard non-PRE bursts than in PRE
and soft non-PRE bursts \cite[Figure \ref{fig: histogram}; see also]
[]{van der Klis90}. In our sample
we have $\sim$ 130 soft non-PRE bursts and $\sim$ 90 hard non-PRE
bursts, but in panel (a) of Figure \ref{fig: histogram}, non-PRE bursts
have more data points than soft non-PRE bursts. (Remember that each point
in that Figure represents a measurement every 0.5 s in the cooling phase
of the bursts.) This means that hard non-PRE bursts have longer decay
times than PRE and soft non-PRE bursts.
The difference of $T_{\rm bb}$
and burst decay time can also be explained by different chemical
composition of the fuel layer \citep{galloway}: If PRE and soft non-PRE
bursts ignite in a He-rich environment, most of the fuel is burnt in a
very short timescale, and therefore these bursts show fast cooling
at the decay phase. The hard non-PRE bursts ignite in a H-rich
environment. Hydrogen burning proceeds more slowly, because it is
limited by the $\beta$-decay moderated by the weak force. During the
cooling of hard non-PRE bursts, the Hydrogen left in the atmosphere 
keeps burning. Also, since the hard non-PRE
bursts ignite in a H-rich environment, they have larger $f_{\rm c}$ than
the PRE and soft non-PRE bursts that ignite in a He-rich environment
\citep{Madej04, Suleimanov, Guver10}, therefore the hard non-PRE bursts
show higher average $T_{\rm bb}$ than the PRE and soft non-PRE bursts. 

We note that the differences in the distribution of $T_{\rm bb}$
are more significant at low flux levels. These differences
can also be due in part to differences in the underlying emission 
of the neutron star due to
continued accretion during the burst. Van Paradijs \& Lewin (1986) argued that
the net burst spectrum is not a blackbody, and at the end of the burst
the $T_{\rm bb}$ of the net burst spectrum is directly related to the
neutron-star surface temperature just before the burst. Therefore,
differences in the average $T_{\rm bb}$ may reflect differences in the
neutron-star surface temperature, depending on the state of the source
at the time the burst starts. However, there are significant differences
in the distribution of $T_{\rm bb}$ at  low flux levels between PRE and
soft non-PRE bursts, whereas both types of bursts occur when 4U 1636--53
is in the soft state, and therefore the persistent spectrum before the
bursts should be similar. Also, both types of bursts should ignite in a
layer with similar chemical composition, which makes it difficult to
explain the differences by differences in the pre-burst persistent
emission,  $f_{\rm c}$ or emission areas.

\section{Conclusions}

We studied the cooling phase of type-I X-ray bursts in the LMXB 4U
1636--53 using all available RXTE data. We divided the bursts in three
groups according to their properties and the spectral state of the
source at the time the burst started: Photospheric radius expansion
(PRE), hard and soft non-PRE bursts, respectively. (Soft non-PRE and PRE
bursts occurred in the soft state of the source.) We found that, during
the cooling phase of the bursts:

\begin{itemize}

\item For all types of bursts, the average bolometric-flux vs.
temperature relation, $F_{\rm b} - T_{\rm bb}$, is significantly different than the $F_{\rm b}
\propto T^4_{\rm bb}$ relation that would be expected if the apparent emitting
area on the neutron star remained constant during the decay of the
bursts. 

\item For all types of bursts, the average $F_{\rm b} - T_{\rm bb}$ relation cannot be
described by single power law over the whole flux range.

\item The average $F_{\rm b} - T_{\rm bb}$ relation is significantly different for PRE,
hard non-PRE and soft non-PRE bursts.

\item In line with the previous conclusions, the temperature
distribution at different flux levels is significantly different for
PRE, hard non-PRE and soft non-PRE bursts.

\end{itemize}

These results imply that either the emitting area on the neutron-star
surface or, most likely, the colour-correction factor changes during the
cooling phase of X-ray bursts.

We calculated the colour-correction factor separately for the three
types of bursts. Compared to the models of \cite{Suleimanov}, 
the dependence of the colour-correction factor with luminosity
(in Eddington units) is consistent with a scenario in which the main
source of fuel in hard non-PRE bursts is hydrogen, whereas for soft
non-PRE and PRE bursts the main source of fuel is helium. 

Based on the dependence of the colour-correction factor with luminosity
at high luminosity, we suggest that at the beginning of the cooling
phase of the bursts there is an enhanced metal abundance in the neutron
star atmosphere, and that the relative metal abundance decreases as the
burst flux decreases.

\section*{Acknowledgments}

This research has made use of data obtained from the High Energy
Astrophysics Science Archive Research Center (HEASARC), provided by
NASA's Goddard Space Flight Center. We thank Jean in't Zand,  Andrew Cumming,
Juri Poutanen, Tod Strohmayer, Duncan Galloway and Valery Suleimanov 
for useful discussions. GZ acknowledges useful discussions with the participants 
of the Lorentz Center workshop "X-ray bursts and burst oscillations". 


\end{document}